\documentclass[final]{article}

\usepackage{arxiv}

\usepackage[utf8]{inputenc} 
\usepackage[T1]{fontenc}    
\usepackage{hyperref}       
\usepackage{url}            
\usepackage{booktabs}       
\usepackage{amsfonts}       
\usepackage{nicefrac}       
\usepackage{microtype}      
\usepackage{lipsum}
\usepackage{graphicx}

\usepackage{xcolor}
\usepackage{tikz}
\usepackage{amsmath}
\usepackage{amssymb}

\usepackage{lineno}
\usepackage{algorithm2e}
\usepackage{textcomp} 
\usepackage{subcaption}
\usepackage{adjustbox}
\usepackage{booktabs}

\modulolinenumbers[5]
\usetikzlibrary{calc}
\usetikzlibrary{intersections}
\usetikzlibrary{through}

\title{Explicit Basis Function Kernel Methods for Cloud Segmentation in Infrared Sky Images}

\author{
 Guillermo Terr\'en-Serrano \\
  Department of Electrical and Computer Engineering \\
  The University of New Mexico \\
  Albuquerque, NM 87131, United States\\
  \texttt{guillermoterren@unm.edu} \\
 \And
  Manel Mart\'inez-Ram\'on \\
  Department of Electrical and Computer Engineering \\
  The University of New Mexico \\
  Albuquerque, NM 87131, United States\\
  \texttt{manel@unm.edu} \\
}

\begin{document}

\maketitle

\begin{abstract}
    Photovoltaic systems are sensitive to cloud shadow projection, which needs to be forecasted to reduce the noise impacting the intra-hour forecast of global solar irradiance. We present a comparison between different kernel discriminative models for cloud detection. The models are solved in the primal formulation to make them feasible in real-time applications. The performances are compared using the j-statistic. The infrared cloud images have been preprocessed to remove debris, which increases the performance of the analyzed methods. The use of the pixels’ neighboring features also leads to a performance improvement. Discriminative models solved in the primal yield a dramatically lower computing time along with high performance in the segmentation.
\end{abstract}


\section{Introduction}

Moving clouds produce energy generation interruptions in Photovoltaic (PV) systems, often causing them to be out of the grid operator’s admissible range \cite{LAPPALAINEN2017}. To improve the overall performance of PV systems, cloud dynamics information must be included in solar irradiance forecast \cite{FURLAN2012}.

Cloud forecasting is also used for home appliance management in events when the solar irradiance available at the time of generation is severely attenuated \cite{Cheng2017b, CHOU2019, TERREN2020b}. For smart grids that require a intra-hour Global Solar Irradiance (GSI) forecast, ground based sky images must be used \cite{KONG2020}. Visible light cameras present difficulties in practice because of the saturation in the circumsolar region \cite{Cheng2017a, Fu2013, Shi2019, Taravat2015}. Some authors have used structures to block the Sun's direct irradiance, but they partially obstruct the images \cite{Chow2011, Dev2017, Li2015, Ye2019}, decreasing the forecasting performance \cite{Yang2018}. Infrared (IR) imaging \cite{MAMMOLI2019} allows the derivation of physical features of the clouds such as temperature \cite{Escrig2013} and altitude, which can in turn be used to model physical processes of the clouds. Radiometric measures \cite{SHAW2005, SHAW2013} allow the extraction of statistical cloud features \cite{THURAIRAJAH2007}.

Real time cloud segmentation using kernel learning is problematic, due to the large quantity of pixels in an image to be classified and the even larger amount of training samples used. The training involves the use of a matrix of dimensions $N \times N$, where $N$ is the number of training samples \cite{Zhuo2014} that have a computational time of $\mathcal{O} (N^3)$. The testing time includes the manipulation of a matrix of $N \times M$ elements, where $M$ is the number of test samples (with the exception of the Support Vector Machine, where the testing time can be reduced due to the sparse nature of its solutions).

In this paper we present a methodology for cloud segmentation in IR images. The sky imager is mounted on a solar tracker. The IR camera records thermal images that allow physical features of clouds to be extracted for segmentation. Image prepossessing \cite{TERREN2020a} and feature extraction are used prior to the cloud classification using kernel methods. In particular, we compare Ridge Regression (RRC), Support Vector Classification (SVC), and Gaussian Processes for Classification (GPC). The methods are implemented in a primal formulation using explicit basis functions that map the data into a finite dimensional Hilbert space through Volterra’s expansions. The training and testing time is dramatically reduced with the respect to the dual formulation. The primal formulation requires manipulations of a $n \times n$ dimensions matrix, where $n$ is the dimension of the Hilbert space and $n << N$.

\section{Feature Extraction}

The IR images are processed before the features are extracted. Image processing is performed in two steps. First, the germanium outdoor lens irradiance scattering effect is removed. This effect is produced by debris such as dust particles and water stains. Second, the atmospheric irradiance effect is removed. The atmospheric effect is composed of the Sun’s direct irradiance and the scattered irradiance from the atmosphere. Two independent models are applied to remove these effects \cite{TERREN2020a}.

\subsection{Infrared Images}

A pixel of the camera frame is defined by a pair of Euclidean coordinates  $i,j$. The height of a particle in the Troposphere can be approximated using the Moist Adiabatic Lapse Rate function \cite{Hess1959, Stone1979} The temperature of the pixels in an image from the radiometric IR camera are $\mathbf{T} = \{ T_{i,j} \in \mathbb{R}^+ \mid \forall i = 1, \ldots, M, \ \forall j = 1, \ldots, N \}$ in centi-Kelvin. Knowing the temperature of the pixels, their heights are computed and defined as $\mathbf{H} = \{ H_{i,j} \in \mathbb{R}^+ \mid \forall i = 1, \ldots, M, \ \forall j = 1, \ldots, N \}$ in kilometers. The temperature of the pixels, after removing debris, are defined as $\mathbf{T}^\prime = \{ T^\prime_{i,j} \in \mathbb{R}^+ \mid \ \forall i = 1, \ldots, M, \ \forall j = 1, \ldots, N \}$, and their heights are $\mathbf{H}^{\prime} = \{ H_{i,j}^\prime \in \mathbb{R}^+ \mid \forall i = 1, \ldots, M, \ \forall j = 1, \ldots, N \}$. After removing both the germanium outdoor lens effect and the atmospheric effect from the images, temperature differences of the pixels with respect to the tropopause's temperature are $\mathbf{\Delta T} = \{ \Delta T_{i,j} \in \mathbb{R} \mid \forall i = 1, \ldots, M, \ \forall j = 1, \ldots, N \}$, and their heights are $\mathbf{H}^{\prime \prime} = \{ H_{i,j}^{\prime \prime} \in \mathbb{R}^+ \mid \forall i = 1, \ldots, M, \ \forall j = 1, \ldots, N \}$.

The temperature differences are normalized to 8 bits, $\mathbf{I} = \{ i_{i,j} \in \mathbb{N}^{2^{8}} \mid \forall i = 1, \ldots, M, \ \forall j = 1, \ldots, N \}$. The lowest temperature is subtracted and then divided by the maximum feasible temperature of the clouds. The feasible temperature is calculated assuming that temperature linearly decreases $9.8^\circ \mathrm{K/km}$ in the tropopause \cite{Hummel1981}. In the sky imager localization, the average height of the tropopause at $36^\circ$ latitude north is $11.5 \mathrm{km}$ \cite{Pan2011}, and average elevation abovesea level is $1.52\mathrm{km}$.

The weighted Lucas-Kanade method \cite{TERREN2020b, SIMON2003} is implemented to compute the velocity vectors. This method
uses two consecutive images ${\bf I}^{k-1}$, ${\bf I}^{k}$ . The velocity vectors are defined as $\mathbf{V}^k = \{ \mathbf{v}_{i,j} = (u,v)^k_{i,j} \in \mathbb{R}^2 \mid \ \forall i = 1, \ldots, M, \ \forall j = 1, \ldots, N \}$. The upper index $k$ denoting the frame is omitted from the rest of the document.

\subsection{Feature Vectors}

To segment a pixel $i,j$, a feature vector of the pixel is combined with features from neighboring pixels, which are then used in the classification model. The sets of physical features extracted from a pixel are validated to find the optimal set. The first vector is $\mathbf{x}^{1}_{i,j} = \{ T_{i,j}, \ H_{i,j} \}$, the second vector is $\mathbf{x}^{2}_{i,j} = \{ T_{i,j}^\prime, \ H_{i,j}^\prime \}$, the third vector is $\mathbf{x}^{3}_{i,j} = \{ \Delta T_{i,j}, \ H_{i,j}^{\prime \prime} \}$ and the fourth vector is $\mathbf{x}^{4}_{i,j} = \{ \mathrm{mag} (\mathbf{v}_{i,j}), \ i_{i,j}, \ \Delta T_{i,j} \}$.

The neighboring pixels included in the input feature vectors are also validated. The term single pixel is used when no features from neighbors are included. The \emph{$1^{st}$ order neighborhood} includes the features of four pixels closest to pixel $i,j$, and the \emph{$2^{nd}$ order neighborhood} includes the features of eight pixels closest to pixel $i,j$. The vectors of features are:
\begin{itemize}
    \item Single pixel: 
            $\{ \mathbf{x}_{i,j} \}, \quad \forall i,j = i_1,j_1, \ldots, i_{M}, j_{N}$.
    \item $1^{st}$ order neighborhood: $\{\mathbf{x}_{i-1,j}, \ \mathbf{x}_{i,j - 1}, \ \mathbf{x}_{i,j + 1}, \ \mathbf{x}_{i+1,j} \}$.
            
    \item $2^{nd}$ order neighborhood: $\{\mathbf{x}_{i-1,j}, \ \mathbf{x}_{i,j - 1}, \ \mathbf{x}_{i,j + 1}, \ \mathbf{x}_{i+1,j}, \ \mathbf{x}_{i - 1,j -1}, \ \mathbf{x}_{i - 1,j + 1}, \ \mathbf{x}_{i + 1,j + 1}, \ \mathbf{x}_{i + 1,j + 1} \}$.
\end{itemize}

\section{Primal Formulation of Kernel Methods for Classification}

The chosen kernel is a polynomial expansion defined as $\varphi : \mathcal{X} \mapsto \mathcal{P}^n$, where $n$ is the order of the expansion. The dimension of the output space is defined as $\mathcal{P}^n =  ( n + d) ! / n!$, so when the transformation is applied to a covariate vector $\mathbf{x}_i \mapsto \varphi ( \mathbf{x}_i )$, a vector is expanded to the $\mathcal{P}^n$-dimension space $\varphi (\mathbf{x}_i) \in \mathbb{R}^{\mathcal{P}^n}$. The polynomial expansion of the dataset $\mathcal{D} = \{ \mathbf{\Phi}, \ \mathbf{y} \}$, is defined in matrix form as,

\begin{align}
    \mathbf{\Phi} = 
    \begin{bmatrix}
    \varphi \left( \mathbf{x}_1 \right) & \hdots & \varphi \left( \mathbf{x}_N \right)
    \end{bmatrix} \in \mathbb{R}^{ \mathcal{P}^n \times N }, \quad \mathbf{y} = 
    \begin{bmatrix}
    y_1 \\
    \vdots \\
    y_N
    \end{bmatrix}, 
\end{align}
where $y_i \in \{ 0, \ 1\}$ which are labels for a pixel representing clear or cloudy conditions, respectively. The polynomial
expansion used in the primal formulated kernel for RRC, SVC and GPC is defined as,
\begin{align}
\begin{split}
    \varphi \left( {\bf x}_i \right) &= \left[ a_0 ~\cdots~ a_j x_j ~\cdots~ a_{j,k} x_j x_k ~\cdots~ a_{j,k,l} x_j x_k x_l ~\cdots~ \right]^\top \in \mathbb{R}^{\mathcal{P}^n}, \\
    & \quad \quad \quad \forall j,k,l \cdots = 1, \dots, D
\end{split}
\end{align}
where the scalar $a_0, a_j, ~a_{j,k},~a_{j,k,l}, ~\cdots \in \mathbb{R}$ is chosen so that the corresponding dot product in the space can be written
\begin{align}
    \varphi \left( \mathbf{x}_i \right)^\top \varphi \left(\mathbf{x}_i \right) = \left[a_0  + \mathbf{x}_i^{\top} \mathbf{x}_i \right]^n
\end{align}

\subsection{Ridge Regression}

RRC is an optimization problem which aims to find the parameters that minimize the mean squared error. In this model, the parameters $\mathbf{w}$ are regularized using the quadratic norm,
\begin{align}
    \label{eq:ridge_regression}
    \min_{\mathbf{w}} \ \sum_{i = 1}^{N} \left( \mathbf{y} - \mathbf{w}^\top \mathbf{\Phi} \right)^2 + \gamma \| \mathbf{w} \|_2.
\end{align}
where $\gamma$ is the regularization parameter.

This model has a quadratic loss function. Therefore, the optimal parameters  $\bar{\mathbf{w}}$ can be found analytically,
\begin{equation}
    \begin{split}
    0 &= \frac{\partial}{\partial \mathbf{w}} \left[ \left( \mathbf{y} - \mathbf{w}^\top \mathbf{\Phi} \right)^\top \left( \mathbf{y} - \mathbf{w}^\top\mathbf{\Phi} \right) + \gamma \mathbf{tr} \left( \mathbf{w}^\top \mathbf{w} \right) \right] \\
    0 &= 2 \left[ \mathbf{\Phi}\left(  \mathbf{w}^\top \mathbf{\Phi}  - \mathbf{y} \right) + \gamma \mathbf{w} \right]\\
    \bar{\mathbf{w}} &= \left( \mathbf{\Phi} \mathbf{\Phi}^\top + \gamma \mathbf{I}\right)^{-1} \mathbf{\Phi} \mathbf{y}.
    \end{split}
\end{equation}

In this case, as the model is for classification, a sigmoid function is applied to the prediction, 
\begin{align}
    \begin{split}
    p \left( \mathcal{C}_1 \mid \varphi \left( \mathbf{x}_* \right), \mathcal{D} \right) &= \frac{1}{1 + \exp \left( - \mathbf{\bar{w}}^\top \varphi \left( \mathbf{x}_* \right) \right)} \\
    p \left( \mathcal{C}_2 \mid \varphi \left( \mathbf{x}_* \right), \mathcal{D} \right) &= 1 - p \left( \mathcal{C}_1 \mid \varphi \left( \mathbf{x}_* \right), \mathcal{D}  \right).
    \end{split}
\end{align}
The result is probability between 0 and 1. The classification threshold is initially set to 0.5, thus the class with higher probability is the predicted class. Lately, it is explained the method implemented to cross-validate the threshold, so that the different models have the same objective function.

\subsection{Primal solution for Support Vector Machines}

We propose to solve a SVC for binary classification in the primal to limit the complexity of the model for cloud segmentation due to the large number of pixels samples \cite{Navia2001, HSU2010}. For the SVC, the dataset is defined as  $\mathbf{X} = \{ \mathbf{x}_i \in \mathbb{R}^D \mid \forall i = 1, \ldots, N  \}$ and $\mathbf{y} = \{ y_i \in \{ -1,+1 \} \mid \ \forall i = 1, \ldots, N \}$. The primal formulation of the SVC is the following unconstrained optimization problem,
\begin{equation}
        \min _{\mathbf{w}} \frac{1}{2} \|\mathbf{w} \| + C \sum_{i = 1}^N \xi({\bf w};{\bf x}_i,y_i) 
\end{equation}
which is a maximum margin problem problem \cite{FAN2008}. When the $\varepsilon$-loss insensitive is applied to the model, the formulation is
\begin{align}
    \label{eq:linear_svm}
    \min _{\mathbf{w}} \frac{1}{2} \| \mathbf{w}\|_2+C \sum_{i=1}^{N} \left( \max \left[ 0,1-y_i \mathbf{w}^\top \varphi \left( \mathbf{x}_i \right)  \right] \right)^2,
\end{align}
where $C$ is the complexity parameter

The linear SVC do not have a probabilistic output. To transform the output into a probability measure, we use the distance of a sample to the hyper-plane,
\begin{equation}
    \begin{split}
    p \left( \mathcal{C}_1 \mid \varphi \left( \mathbf{x}_* \right), \mathcal{D}  \right) &= \frac{1}{1 + \exp \left( - \mathbf{\bar{w}}^\top \varphi \left( \mathbf{x}_* \right) \right)} \\
    p \left( \mathcal{C}_2 \mid \varphi \left( \mathbf{x}_* \right), \mathcal{D}  \right) &= 1 - p \left( \mathcal{C}_1 \mid \varphi \left( \mathbf{x}_* \right), \mathcal{D} \right),
    \end{split}
\end{equation}
and the sigmoid function (as in RRC).

\subsection{Primal solution for Gaussian Processes}

When a GPC is formulated in the primal, it is commonly known as Bayesian logistic regression \cite{MURPHY2012, Rasmussen2006, Jaakkola1997}. As the GPC does not have an analytical solution, the Laplace approximation is applied to solve this problem,
\begin{align}
    p \left( \mathbf{w} \mid \mathcal{D} \right) \propto  p \left( \mathbf{y} \mid \mathbf{\Phi}, \mathbf{w} \right) \cdot  p \left( \mathbf{w} \right).
\end{align}

The likelihood function $ p ( y_i \mid \mathbf{\Phi}, \mathbf{w} ) = \prod_{i = 1}^N \hat{y}_i^{y_i} ( 1 - \hat{y} )^{1 - y_i}$, where $\mathbf{\hat{y}} = [ \hat{y}_1 \dots \hat{y}_N ]^\top$ are the predictions, is a Bernoulli distribution, and the prior $ p ( \mathbf{w} ) \sim \mathcal{N} \left( \mathbf{w}  \mid \boldsymbol{\mu}_0, \mathbf{\Sigma_0} \right)$ is a Normal distribution. This combination of distributions leads to a posterior that is not Gaussian. Laplace approximation assumes that the posterior is Gaussian$ q ( \mathbf{\bar{w}} ) \sim \mathcal{N} ( \mathbf{w} \mid \mathbf{\bar{w}}, \mathbf{\Sigma_n} )$.

The marginal log-likelihood is maximized to find the optimal parameters $\bar{\mathbf{w}}$,
\begin{equation}
    \begin{split}
    \label{eq:martinal_log_likelihooh}
    \log p \left( \bar{\mathbf{w}} \mid \mathcal{D} \right) &= -\frac{d}{2} \log 2\pi - \frac{1}{2}\log |\mathbf{\Sigma}_0|-\frac{1}{2} \left( \mathbf{w} - \boldsymbol{\mu}_0 \right)^\top \mathbf{\Sigma}_0^{-1} \left( \mathbf{w} - \boldsymbol{\mu}_0 \right) + \\
    & + \sum_{i = 1}^N \left[ y_i \log \hat{y}_i + \left( 1 - y_i \right) \log \left( 1 - \hat{y}_i \right) \right],
    \end{split}
\end{equation}
where $\hat{y}_i = \sigma ( \mathbf{w}^\top \varphi ( \mathbf{x}_i ) ) = 1 / [ 1 + \exp ( - \mathbf{w}^\top \varphi ( \mathbf{x}_i ) )]$, is the sigmoid function.

The covariance of the posterior distribution is found analytically as the inverted Hessian of the negative log-posterior,
\begin{align}
    \mathbf{\Sigma}_n^{-1} = \mathbf{\Sigma}_0^{-1} + \sum_{i = 1}^N y_i \left( 1 - y_i \right) \varphi \left( \mathbf{x}_i \right) \varphi \left( \mathbf{x}_i\right)^\top.
\end{align}

As the convolution of a sigmoid function with a Normal distribution is intractable, the sigmoid function is approximated by a probit function. The approximated predictive distribution is,
\begin{equation}
    \begin{split}
    p \left( C_1 \mid \mathcal{D}  \right)
    &= \int \sigma \left( \boldsymbol{\alpha}  \right)\cdot \mathcal{N} \left( \boldsymbol{\alpha}  \mid \boldsymbol{\mu_\alpha}, \boldsymbol{\sigma}_\alpha \right) d \boldsymbol{\alpha} \\
    &\approx \sigma \left( \Phi \left( \boldsymbol{\sigma}_\alpha \right) \cdot \boldsymbol{\mu}_\alpha \right),
    \end{split}
\end{equation}
where $\boldsymbol{\alpha} = \mathbf{\bar{w}}^\top \mathbf{\Phi}$, the predictive mean is $\boldsymbol{\mu}_\alpha = \mathbf{\bar{w}}^\top \mathbf{\Phi}$, and the variance is $\boldsymbol{\sigma}^2_\alpha = \mathbf{\Phi}^\top \mathbf{\Sigma}_n \mathbf{\Phi}$. The probit approximation of a sigmoid is $\Phi ( \boldsymbol{\sigma}_\alpha ) = ( 1 + \boldsymbol{\sigma}_\alpha \pi/8 )^{-1/2}$. Once the probability of class $\mathcal{C}_1$ is computed
using Eq. (12), the probability of class $\mathcal{C}_2$ is,
\begin{align}
    p \left( C_2 \mid \varphi \left( \mathbf{x_*} \right), \mathcal{D}  \right)
    &= 1 - p \left( C_1 \mid \varphi \left( \mathbf{x_*} \right), \mathcal{D} \right).
\end{align}

\section{J-Statistic}

The Youden’s j-statistic is a statistical test that evaluates the performances of a dichotomous classification \cite{YOUDEN1950},
\begin{equation}
    j = sensitivity - specificity - 1.
\end{equation}

In a dichotomous classification, the probabilities of each class are $p\left(\mathcal{D}\mid\mathcal{C}_1\right) = p\left(\mathcal{C}_1\mid\mathcal{D}\right) \lambda$ and $p\left(\mathcal{D}\mid\mathcal{C}_2\right)= 1 - p\left(\mathcal{D}\mid\mathcal{C}_1 \right)$. The optimal j-statistic is computed at each point of the Receiver Operating Characteristic (ROC) \cite{FAWCETT2006}, cross-validating the virtual prior $\lambda$. The optimal value of $\lambda$ is that which produces the best j-statistic in the ROC curve,
\begin{equation}
    \begin{split}
    \label{eq:virtual_prior}
    p \left( \mathcal{D} \mid \mathcal{C}_k \right) &= \frac{p \left( \mathcal{C}_k \mid \mathcal{D} \right) p \left( \mathcal{C}_k  \right)}{p \left( \mathcal{D} \right)} \\ 
    &\propto p \left( \mathcal{C}_k \mid \mathcal{D}  \right) p \left( \mathcal{C}_k \right) \\
    &\propto p \left( \mathcal{C}_k \mid \mathcal{D}  \right) \lambda.
    \end{split}
\end{equation}

A class $\mathcal{C}_k$ is assigned to a new sample following the MAP criteria,
\begin{align}
    \hat{y}_* = \underset{k}{\operatorname{argmax}} \ p \left( \mathcal{C}_k \mid \mathbf{x}_*, \mathcal{D} \right) \lambda.
\end{align}

\section{Experiments}

The proposed segmentation methods utilize data acquired by a sky imager equipped with a solar tracker that updates its pan and tilt every second, maintaining the Sun in a central position in the images throughout the day. The sensor is a Lepton1 radiometric IR camera sensitive to wavelengths from 8 to 14 $\mu m$. The intensity of the pixels in a frame are measurements of temperature in centi-kelvin degrees. The resolution of an IR image is $80 \times 60$ pixels. The diagonal field of view of the camera is $60^\circ$. The sky imager is localized on the roof of the UNM-ME building in Albuquerque, NM. The data is accessible in a Dryad repository \cite{GIRASOL}.

\begin{figure}[!ht]
    \begin{subfigure}{0.325\linewidth}
        \includegraphics[scale = 0.325]{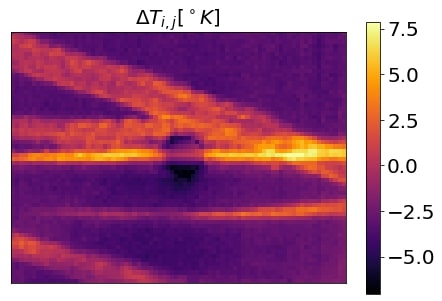}
    \end{subfigure}
    \begin{subfigure}{0.325\linewidth}
        \includegraphics[scale = 0.325]{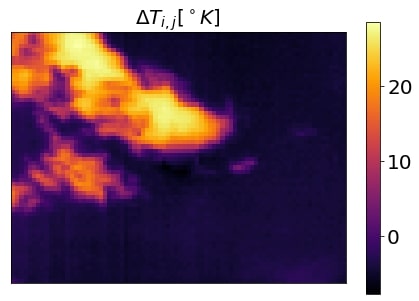}
    \end{subfigure}
    \begin{subfigure}{0.325\linewidth}
        \includegraphics[scale = 0.325]{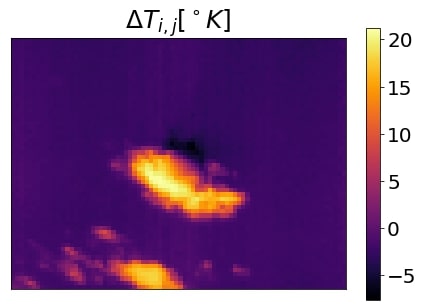}
    \end{subfigure}
    \begin{subfigure}{0.325\linewidth}
        \includegraphics[scale = 0.325]{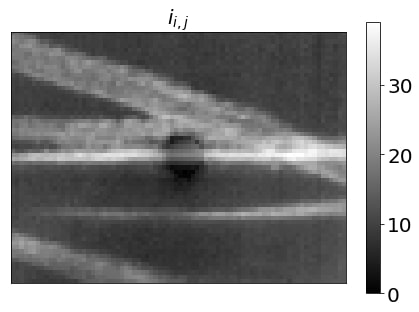}
    \end{subfigure}
    \begin{subfigure}{0.325\linewidth}
        \includegraphics[scale = 0.325]{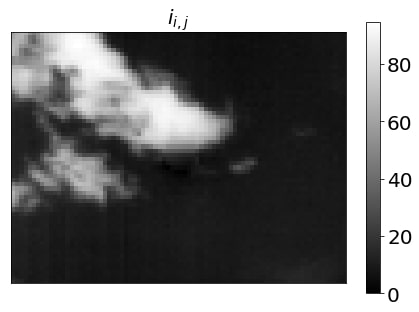}
    \end{subfigure}
    \begin{subfigure}{0.325\linewidth}
        \includegraphics[scale = 0.325]{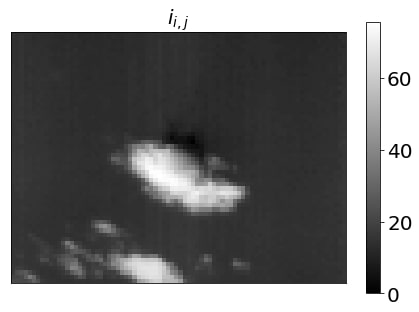}
    \end{subfigure}
    \begin{subfigure}{0.325\linewidth}
        \includegraphics[scale = 0.325]{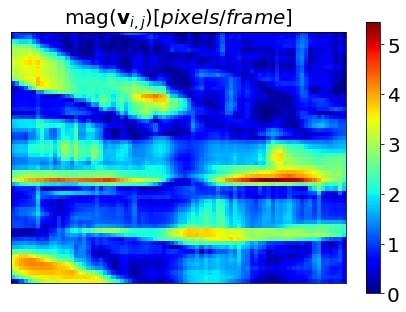}
    \end{subfigure}
    \begin{subfigure}{0.325\linewidth}
        \includegraphics[scale = 0.325]{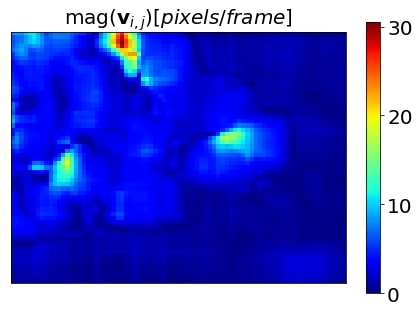}
    \end{subfigure}
    \begin{subfigure}{0.325\linewidth}
        \includegraphics[scale = 0.325]{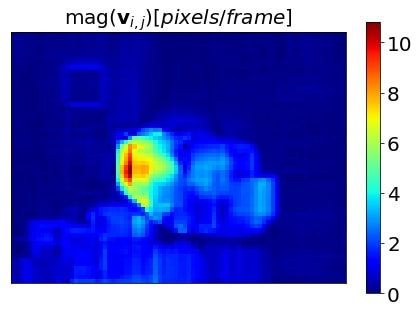}
    \end{subfigure}
    \begin{subfigure}{0.325\linewidth}
        \includegraphics[scale = 0.325]{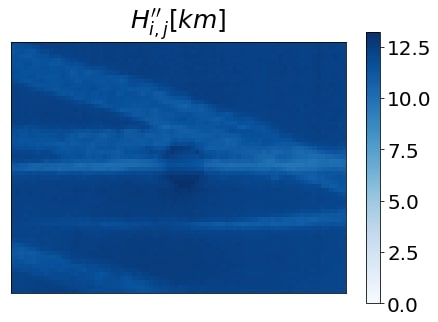}
    \end{subfigure}
    \begin{subfigure}{0.325\linewidth}
        \includegraphics[scale = 0.325]{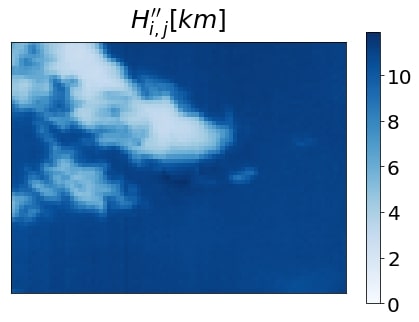}
    \end{subfigure}
    \begin{subfigure}{0.325\linewidth}
        \includegraphics[scale = 0.325]{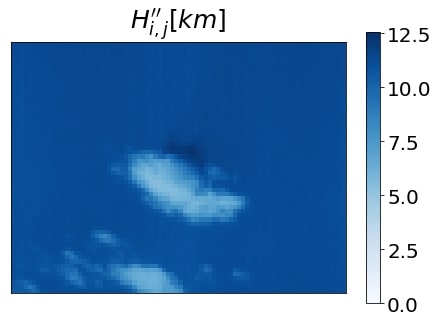}
    \end{subfigure}
\caption{Physical features extracted from the three testing IR images after removing the cyclo-stationary processes. The first row shows the incremental temperatures of a pixel with respect to the tropopause in centi-kelvin degrees. The second row shows the normalized intensities used to compute the velocity vectors. The third row shows the velocity vectors computed using the weighted Lucas-Kanade method. The fourth row shows the heights of the pixels computed using the Moist Adiabatic Lapse Rate}
\label{fig:features}
\end{figure}

The database contains samples of consecutive IR images recorded every 15 seconds between sunrise and sunset. The samples were recorded in 2017, 2018 and 2019. The IR images are first processed to extract the features included in the feature vectors (see Fig. 1). Twelve IR image were selected to cross-validate the parameters of the cloud segmentation models. To develop a segmentation model that performs accurate classifications with different sky conditions and cloud types, the dataset of selected images includes images acquired at different times of the day in different seasons with clear-sky, partially cloudy or covered sky conditions. The types of clouds included in the dataset are stratocumulus, cumulus, cirrocumulus, altocumulus, nimbus, contrail and altostratus.

The pixels in the selected IR images that form the dataset were manually labelled as clear $y_{i,j}=0$ or cloudy $y_{i,j}=1$. The dataset has a total of 57,600 pixels. The background temperature in the IR images is the temperature of the tropopause. This temperature was used to distinguish which pixels match the background temperature or have a higher temperature (which indicates the presence of clouds). The samples are chronologically divided to form the training (earlier dates) and testing dataset (later dates). The training dataset has 7 samples, and the testing dataset has 5 samples. Both datasets have samples of clear-sky, partially cloudy and covered sky. The training set has a total of 33,600 pixels and the testing set has 24,000 pixels with their respective labels.

The cross-validation of the parameters is performed implementing the Leave-One-Out (LOO) method. The LOO iterates the training set, leaving out one of the samples for validation at a time, and the model is trained with the remaining samples. The j-statistic is computed for each validation sample. The best average j-statistic obtained in all LOO validations is proposed as the model selection criteria. In this way, a model is fitted using the training set for each set of parameters and a virtual prior $\lambda$ in Eq. (17). The virtual prior is adjusted to the optimal j-statistic using the predicted probabilities of each class in each combination of parameters that is cross-validated. The virtual prior $\lambda$ is cross-validated in all the models. In addition to the virtual prior $\lambda$, each model has parameters which have to be cross- validated. The RRC have the regularization $\gamma$ parameter in Eq. (4). The SVC has the complexity term $C$ of the loss function in Eq. (8). The prior mean $\boldsymbol{\mu}_0$ and covariance matrix $\boldsymbol{\Sigma}_0$ in Eq. (11) are GPC hyperparameters. The hyperparameters are simplified to $\boldsymbol{\mu}_0 \triangleq \mathbf{0}$ and $\boldsymbol{\Sigma}_0 \triangleq \gamma \mathbf{I}$, so only the parameter $\gamma$ of the covariance matrix is cross-validated.

\begin{table}[htb!]
    \centering
    \small
    \setlength{\tabcolsep}{12.pt} 
    \renewcommand{\arraystretch}{1.} 
    \begin{tabular}{lcccccc}
        \toprule
        {} & \multicolumn{3}{c}{J-statistic [\%]} & \multicolumn{3}{c}{Time [ms]} \\
        Feature Vector & Single & $1^{st}$ Order& $2^{nd}$ Order&  Single & $1^{st}$ Order & $2^{nd}$ Order\\
        \midrule
        \multicolumn{7}{c}{RRC} \\
        \midrule
        $\mathbf{x}^1_{i,j}$ & 51.74 & 48.32 & 47.94 & 5.36 & 3.07 & 2.40 \\
        $\mathbf{x}^2_{i,j}$ & 60.38 & 55.64 & 55.84 & 3.18 & 3.06 & 2.10 \\
        $\mathbf{x}^3_{i,j}$ & 80.97 & 82.83 & 82.75 & 4.22 & 5.31 & 2.48 \\
        $\mathbf{x}^4_{i,j}$ & 88.92 & \textbf{91.64} & 89.04 & 1.90 & \textbf{2.15} & 0.91 \\
        $\mathcal{P}^2(\mathbf{x}^1_{i,j})$ & 47.09 & 46.02 & 42.72 & 4.54 & 2.20 & 1.28 \\
        $\mathcal{P}^2(\mathbf{x}^2_{i,j})$ & 45.96 & 30.13 & 33.43 & 3.26 & 0.95 & 1.36 \\
        $\mathcal{P}^2(\mathbf{x}^3_{i,j})$ & 79.56 & 88.57 & 90.42 & 2.77 & 1.10 & 1.12 \\
        $\mathcal{P}^2(\mathbf{x}^4_{i,j})$ & 49.08 & 59.14 & 67.60 & 3.50 & 1.23 & 2.28 \\
        \midrule
        \multicolumn{7}{c}{SVC} \\
        \midrule
        $\mathbf{x}^1_{i,j}$ & 47.70 & 48.81 & 48.55 & 5.25 & 0.52 & 0.50 \\
        $\mathbf{x}^2_{i,j}$ & 44.53 & 44.31 & 52.35 & 3.66 & 1.09 & 0.47 \\
        $\mathbf{x}^3_{i,j}$ & 72.12 & 72.58 & 69.48 & 4.54 & 2.57 & 1.03 \\
        $\mathbf{x}^4_{i,j}$ & \textbf{91.82} & 91.14 & 91.36 & \textbf{3.66} & 1.18 & 1.02 \\
        $\mathcal{P}^2(\mathbf{x}^1_{i,j})$ & 42.65 & 43.46 & 45.49 & 0.40 & 0.75 & 4.40 \\
        $\mathcal{P}^2(\mathbf{x}^2_{i,j})$ & 48.06 & 34.60 & 42.61 & 0.45 & 1.77 & 1.82 \\
        $\mathcal{P}^2(\mathbf{x}^3_{i,j})$ & 73.49 & 75.70 & 76.38 & 2.89 & 1.73 & 4.46 \\
        $\mathcal{P}^2(\mathbf{x}^4_{i,j})$ & 59.23 & 66.04 & 67.96 & 0.46 & 3.41 & 8.18 \\
        \midrule
        \multicolumn{7}{c}{GPC} \\
        \midrule
        $\mathbf{x}^1_{i,j}$ & 8.49 & 45.19 & 45.24 & 75.68 & 79.33 & 107.13 \\
        $\mathbf{x}^2_{i,j}$ & 53.77 & 45.95 & 54.43 & 65.46 & 93.91 & 105.07 \\
        $\mathbf{x}^3_{i,j}$ & 72.39 & 84.00 & 83.40 & 103.05 & 94.57 & 106.69 \\
        $\mathbf{x}^4_{i,j}$ & \textbf{92.41} & 91.45 & 90.70 & \textbf{77.02} & 104.33 & 125.47 \\
        $\mathcal{P}^2(\mathbf{x}^1_{i,j})$ & 36.11 & 40.98 & 43.75 & 69.96 & 152.06 & 520.79 \\
        $\mathcal{P}^2(\mathbf{x}^2_{i,j})$ & 51.13 & 41.22 & 38.97 & 79.99 & 219.08 & 411.87 \\
        $\mathcal{P}^2(\mathbf{x}^3_{i,j})$ & 82.87 & 77.99 & 78.03 & 86.37 & 205.60 & 413.50 \\
        $\mathcal{P}^2(\mathbf{x}^4_{i,j})$ & 73.21 &  79.77 & 83.93 & 93.95 & 312.57 & 975.83 \\
        \bottomrule
    \end{tabular}
    \caption{J-statistic and average testing time obtained in test by the RRC, SVC and GPC models. Groups of three columns show the results by neighborhood of the pixels: single pixel, $1^{st}$ order or $2^{nd}$. The different feature vectors are in rows and when polynomial expansion is applied the vector is denoted as $\mathcal{P}^2(\cdot)$.}
    \label{tab:rrc_table}
\end{table}


The experiments were carried out in the Wheeler high performance computer of UNM-CARC, which uses a SGI AltixXE Xeon X5550 at 2.67GHz with 6 GB of RAM memory per core, 8 cores per node, 304 nodes total, and runs at 25 peak FLOPS (theoretical) in TFLOPS. It has Linux CentOS 7 installed.

\section{Discussion}

RRC, SVC and GPC were solved in the primal formulation, so their performances are feasible for real-time cloud segmentation. The performances of the best models are compared in terms of j-statistic vs. average computing time. The j-statistic and the computing time are evaluated using the images in the test subset.

\begin{figure}[!htb]
    \begin{subfigure}{0.325\linewidth}
        \includegraphics[scale = 0.4]{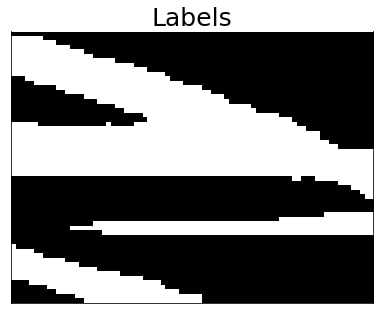}
    \end{subfigure}
    \begin{subfigure}{0.325\linewidth}
        \includegraphics[scale = 0.4]{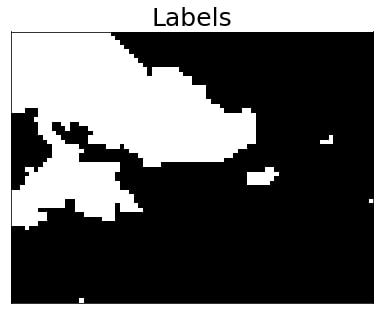}
    \end{subfigure}
    \begin{subfigure}{0.325\linewidth}
        \includegraphics[scale = 0.4]{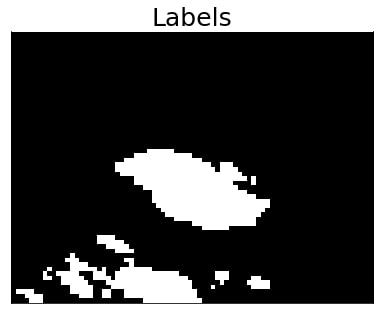}
    \end{subfigure}
    \begin{subfigure}{0.325\linewidth}
        \includegraphics[scale = 0.4]{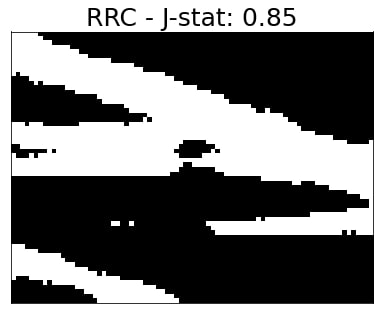}
    \end{subfigure}
    \begin{subfigure}{0.325\linewidth}
        \includegraphics[scale = 0.4]{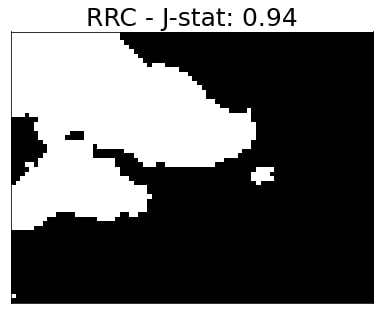}
    \end{subfigure}
    \begin{subfigure}{0.325\linewidth}
        \includegraphics[scale = 0.4]{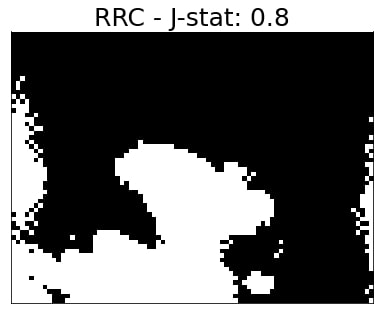}
    \end{subfigure}
    \begin{subfigure}{0.325\linewidth}
        \includegraphics[scale = 0.4]{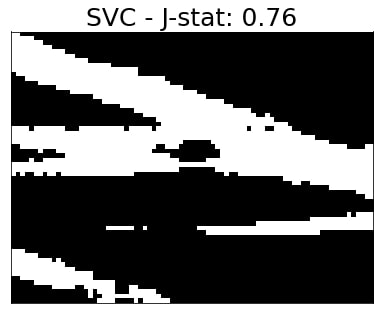}
    \end{subfigure}
    \begin{subfigure}{0.325\linewidth}
        \includegraphics[scale = 0.4]{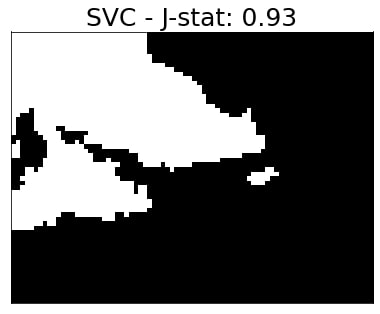}
    \end{subfigure}
    \begin{subfigure}{0.325\linewidth}
        \includegraphics[scale = 0.4]{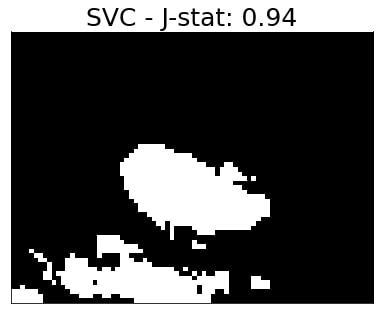}
    \end{subfigure}
    \begin{subfigure}{0.325\linewidth}
        \includegraphics[scale = 0.4]{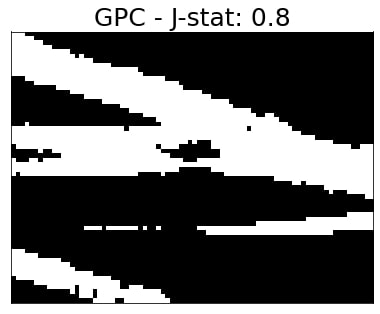}
    \end{subfigure}
    \begin{subfigure}{0.325\linewidth}
        \includegraphics[scale = 0.4]{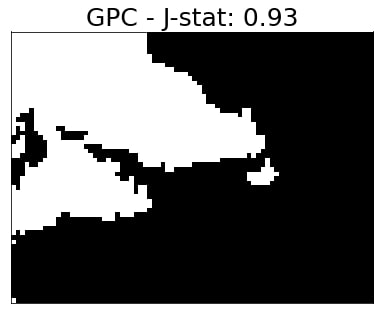}
    \end{subfigure}
    \begin{subfigure}{0.325\linewidth}
        \includegraphics[scale = 0.4]{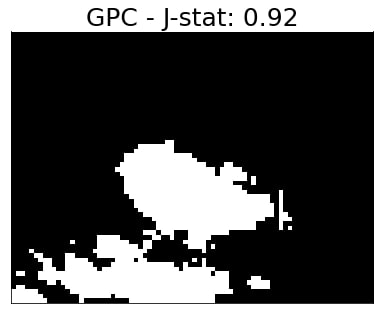}
    \end{subfigure}
\caption{Segmentation of clouds in three testing images. The first row shows the manually segmented images (a pixel containing a cloud is white, a clear-sky pixel is black). The images were segmented using the best RRC model in the second row, the best SVC model in the third row and the best GPC in the fourth row. The RRC achieved the best j-statistic when segmenting the first and second images. The SVC obtained the best j- statistic segmenting the third image. The best average j-statistic was achieved by the best GPC model.}
\label{fig:discriminative_test}
\end{figure}

Fig. 2 shows the discriminative models' j-statistics. The polynomial expansion yields over-fitting. The best model was the linear GPC. The feature vector was $\mathbf{x}^4$ of a single pixel. When the raw temperature and height are used, all models have poor performance. After processing the images with the window model and the atmospheric model, the performances of all methods are increasing but they are still low, ranging between 72.58\% and 84\%. When velocity vectors are added to the features, the methods have higher performances with computational times of 2.2 ms (RRC), 3.7 ms (SVC) and 77 ms (GPC). The best trade-off is SVC, which is 20 times faster than GP with a small difference in accuracy. The image preprocessing and feature extraction time is 0.1 ms for $\mathbf{x}^1$, 4.7 ms for $\mathbf{x}^2$, 99.9 ms for $\mathbf{x}^3$ and 1079 ms for $\mathbf{x}^4$. When preprocessing time is added to the segmentation time the average time required by the models are 1081 ms (RRC), 1083 ms (SVC) and 1156 ms (GPC).

The best j-statistic is achieved by the GPC, but with the largest average testing computing time. If we apply the Pareto front criteria to select a model \cite{BLASCO2008}, RRC and SVC are the most suitable methods. They require low training time, and they are capable of performing accurate and fast segmentation.

\section{Conclusion}

This investigation aims to find an optimal discriminative model for real-time segmentation of clouds in infrared sky images. The performances of the primal formulation of the ridge regression, support vector classifier and Gaussian process for classification are compared. The explicit basis function implemented is a Volterra’s polynomial expansion. The infrared images were processed to remove cyclo-stationary effects and to extract physical features form the clouds. Image processing increases the classification performances with low computational cost. The j-statistic is the metric used to compare the classification performances of the models. The primal formulation of the kernel models results in segmentation models feasible in real-time. The processing of the images to remove the cyclo-stationary effects increases the classification performances of the models. When the feature vector includes features extracted from neighboring pixels, the segmentation performances increased for some models.

Further research will compare the performances of discriminative models with generative models. The generative models will be trained using supervised and unsupervised learning algorithms. The covariance matrix of these models can be simplified to shorten the training and testing times. Markov Random Fields, a class of generative models, include information of the classification of neighboring pixels as the prior. The segmentation algorithm will be improved to detect clouds that are moving in wind flows at different heights. This will be included in a very short- term global solar irradiance forecasting algorithm to increase overall forecasting performances.

\section{Acknowledgments}

This work has been supported by NSF EPSCoR grant number OIA-1757207 and the King Felipe VI endowed Chair. Authors would like to thank the UNM Center for Advanced Research Computing, supported in part by the National Science Foundation, for providing the high performances computing and large-scale storage resources used in this work.

\bibliographystyle{unsrt}  
\bibliography{mybibfile}

\end{document}